# Revealing Atomic-Scale Switching Pathways in van der Waals Ferroelectrics

Xinyan Li[1,2,12], Kenna Ashen[3,12], Chuqiao Shi[1], Nannan Mao[4,5], Saagar Kolachina[3], Kaiwen Yang[6], Tianyi Zhang[4], Sajid Husain[7], Ramamoorthy Ramesh[1,2,7,8], Jing Kong[4], Xiaofeng Qian[3,9,10], Yimo Han[1,2,11,*]

[1]Department of Materials Science and NanoEngineering, Rice University, Houston, TX, USA.
[2]Rice Advanced Materials Institute, Rice University, Houston, TX, USA.
[3]Department of Materials Science and Engineering, Texas A&M University, College Station, TX, USA.
[4]Department of Electrical Engineering and Computer Science, Massachusetts Institute of Technology, Cambridge, MA, USA.
[5]Department of Chemical Engineering, Massachusetts Institute of Technology, Cambridge, MA, USA.
[6]Department of Chemistry, Rice University, Houston, TX, USA
[7]Department of Materials Science and Engineering, University of California, Berkeley, CA, USA.
[8]Department of Physics & Astronomy, Rice University, Houston, TX, USA.
[9]Department of Electrical and Computer Engineering, Texas A&M University, College Station, TX, USA.
[10]Department of Physics and Astronomy, Texas A&M University, College Station, TX, USA.
[11]Smalley-Curl Institute, Rice University, Houston, TX, USA.

[12]These authors contributed equally: Xinyan Li, Kenna Ashen
*Corresponding author: yimo.han@rice.edu (Y. H.)

## Abstract

Two-dimensional van der Waals (vdW) materials hold the potential for ultra-scaled ferroelectric (FE) devices due to their silicon compatibility and robust polarization down to atomic scale. However, the inherently weak vdW interactions enable facile sliding between layers, introducing complexities beyond those encountered in conventional ferroelectric materials and presenting significant challenges in uncovering intricate switching pathways. Here, we combine atomic-resolution imaging under *in-situ* electrical biasing conditions with first-principles calculations to unravel the atomic-scale switching mechanisms in SnSe, a vdW group-IV monochalcogenide. Our results uncover the coexistence of a consecutive 90° switching pathway and a direct 180° switching pathway from antiferroelectric (AFE) to FE order in this vdW system. Atomic-scale investigations and strain analysis reveal that the switching processes simultaneously induce interlayer sliding and compressive strain, while the lattice remains coherent despite the presence of multidomain structures. These findings elucidate vdW ferroelectric switching dynamics at atomic scale and lay the foundation for the rational design of 2D ferroelectric nanodevices.

# Introduction

The broken centrosymmetry in ferroelectric (FE) materials gives rise to microscopic electric dipoles and macroscopic switchable spontaneous polarization, offering enticing opportunities for next-generation non-volatile memories and logic transistors[1-7]. As semiconductor devices continue to scale down, two-dimensional (2D) van der Waals (vdW) ferroelectrics present the potential to maintain ferroelectricity down to atomic thickness while seamlessly integrating with current silicon-based semiconductor technology[8-13]. In 2D vdW materials, such as bilayer hBN[14,15] and transition metal dichalcogenides[16-20], interlayer sliding can break centrosymmetry and induce out-of-plane spontaneous polarization, known as sliding ferroelectrics. However, the electric polarization of these improper ferroelectrics arising from interlayer charge transfer leads to relatively small polarization, posing limitations for practical applications. In contrast, intrinsic 2D vdW proper ferroelectrics exhibit large polarization typically one order of magnitude greater than that of improper sliding ferroelectrics, resulting from the displacement between cations and anions[21-27].

As a representative intrinsic 2D ferroelectric semiconductor, group-IV monochalcogenides (MXs, where M = Ge, Sn, and X = S, Se, Te, such as SnSe) possess robust in-plane ferroelectricity and giant nonlinear optical responses down to monolayer thickness at room temperature[28-32]. Unlike the rigid-like models of 2D sliding ferroelectrics, MXs exhibit intralayer ferroelectric-ferroelastic coupling, where the anisotropic in-plane strain aligns with the polarization direction[28,31,33], introducing tunable parameters such as electric fields and strain[34-37]. Additionally, MX multilayers exhibit intrinsic coupling between interlayer stacking and polarization order, enabling the coexistence of FE and antiferroelectric (AFE) phases with different stacking configurations[35,38-41]. Although several studies have demonstrated reversible polarization switching in MX compounds with non-volatile[36,42-44] and reversible properties[28,33,45-47], the interplay between polarization, strain, and stacking configuration complicates the discovery of energetically favorable switching pathways. This complexity, coupled with the lack of experimental insights, underscores the need for real-time atomic-scale observation of structural evolution during the switching process.

In this study, we combine in-situ atomic-resolution imaging with density functional theory (DFT) calculations to uncover two coexisting, energetically favorable switching pathways in SnSe (anti)ferroelectrics. The *in-situ* biasing scanning transmission electron microscopy (STEM) imaging directly captures the atomic-scale structures of both pathways, including intermediate states and the final FE phases following the switching processes. The results reveal that during the polarization switching process, SnSe undergoes simultaneous interlayer sliding and compressive strain, forming energetically favorable FE states. These findings provide a comprehensive understanding of the switching pathways in SnSe from AFE to FE phases and highlight the intricate interplay between polarization switching, interlayer sliding, and lattice strain in vdW FE systems, which offers valuable insights for optimizing the performance of next-generation semiconductor devices.

## Results

### Energy landscape and switching pathways in SnSe

The atomic model of monolayer SnSe illustrates the relative displacement between Sn and Se ions from the plan view (Figure 1a), which breaks centrosymmetry and generates in-plane spontaneous polarization along the x-axis. Figure 1b displays the side view of monolayer SnSe, corresponding to the armchair (x, long axis) and zigzag (y, short axis) structures. When SnSe is extended to a multilayer system, the stacking configuration is strongly coupled with polarization order between the adjacent layers. The energy landscapes of FE (Figure 1c) and AFE-order (Figure S1a) SnSe with varying stacking configurations show that there are limited numbers of energy favorable states. As shown in the inset of Figure 1c, AA, AB, AC, and AD stackings are defined by the relative interlayer Sn-to-Sn sliding distance (x,y) of (0,0), (0.5$a$,0), (0,0.5$b$), and (0.5$a$,0.5$b$), respectively. Regarding the AFE-order energy landscape and optimized structures (Figure S1a and Table S1), there is only one stable stacking configuration, namely the ground-state (0.3$a$,0) structure. We designate it as AB′ to distinguish it from AB (0.5$a$, 0) stacking. In contrast, FE-order energy landscape identifies metastable AC (0.8 meV/atom) and AB (1.2 meV/atom) stacking configurations, each exhibiting slight differences in polarization (Figure S2). Prior experimental studies confirm that ground-state AB′ AFE and metastable FE structures always coexist in multilayer SnSe flakes[35,38,40]. Therefore, elucidating the key switching pathways from the AFE ground state to metastable FE states in SnSe multilayers is of critical importance.

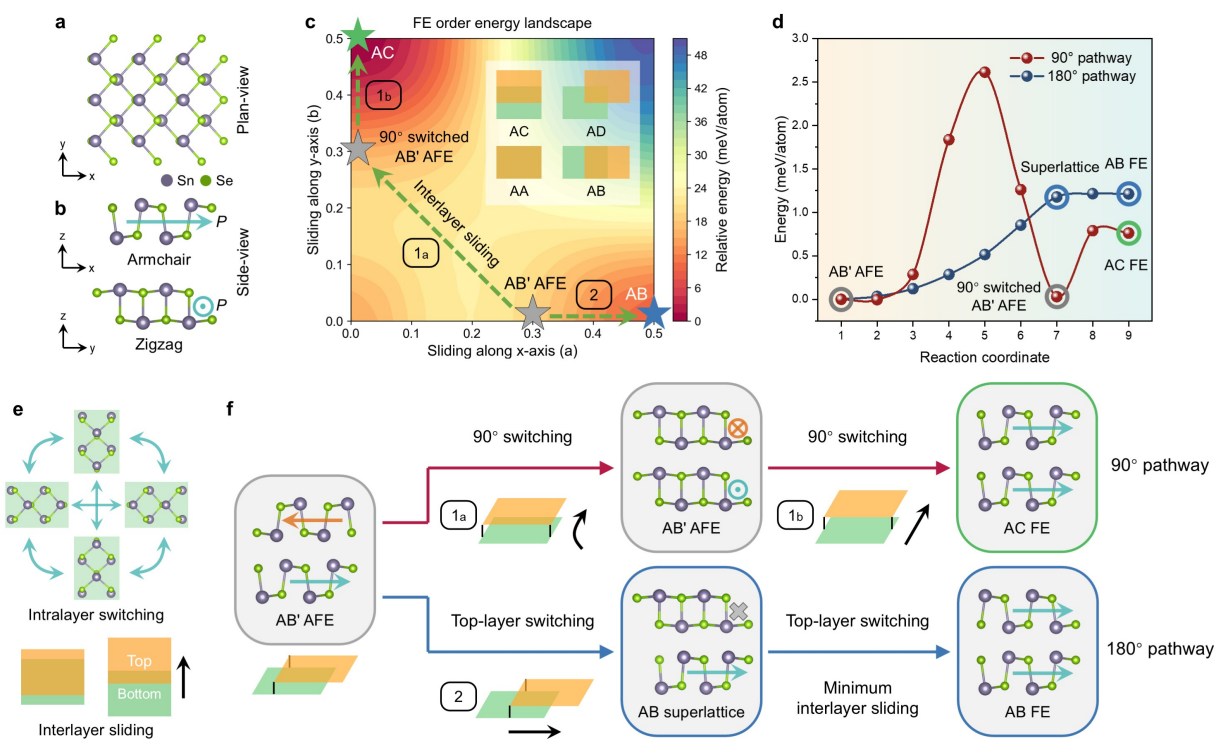

**Figure 1. Complete switching pathways of vdW SnSe.** (a,b) Plan-view (a) and side-view (b) atomic structures of monolayer SnSe. (c) Energy landscape of FE-order multilayer SnSe with different stacking

configurations. The green dashed arrows represent different interlayer sliding pathways from ground-state AFE-order AB′ stacking to FE-order AC or AB stacking energy wells. The orange and green rectangles represent top and bottom layers, respectively. (d) DFT-calculated energy barriers of pathway #1 (90° switching pathway) and pathway #2 (180° switching pathway) corresponding to green dashed arrows in (c). The gray circles correspond to gray stars in (c) (AB′ AFE state). The green/blue circles correspond to green/blue stars in (c) (AC FE/AB FE states). (e) Plan-view schematics of 90° and 180° intralayer polarization switching of electric dipoles within the same layer and interlayer sliding between the adjacent layers. The corresponding side-view schematics of switching are shown in Figure S3. The black arrow represents sliding direction of top layer. (f) Schematics of 90° and 180° switching pathway from AFE-order ground state to two FE-order energy wells through interlayer sliding and intralayer atomic displacement corresponding to green dashed arrows in (c) and energy barriers in (d) The orange cross and blue dot denote polarization vectors pointing out of and into the page, respectively, while the gray cross indicates a nonpolar layer.

Starting from the AB′ AFE ground state, we calculate the switching pathways to AC and AB FE states (green dashed line in Figure 1c) and their corresponding energy barriers (Figure 1d). To simplify the description of the pathways, we define the switching of electric dipoles within a single SnSe layer as "intralayer polarization switching" and the relative displacement between adjacent layers as "interlayer sliding" (Figure 1e and details in Supplementary Note 1).

Along the pathway from AB′ AFE to AC FE, interlayer sliding progresses from $(0.3a,0)$ to $(0,0.5b)$ on the sliding map. Using the solid-state nudged elastic band (SS-NEB) method, we identified that this pathway first undergoes a spontaneous 90° switching to the same AB′ AFE structure (pathway 1a), as confirmed by lattice parameters (Figure S4) and energy values (gray circles in Figure 1d). This is followed by sliding to the AC FE state through a second 90° switching (pathway 1b). It is noted that due to the orthorhombic symmetry of SnSe, the armchair and zigzag directions interchange after the 1a switching pathway (Figure S1 and Supplementary Note 1). The pathway connects two energy wells of equal energy separated by sliding-induced energy barrier (Figure S1b and two gray circles on the red line in Figure 1d). Subsequently, it slides from $(0,0.3b)$ to $(0,0.5b)$ along the 1b pathway, undergoing a second 90° switching to energy favorable AC stacking in FE-order landscape (green circle on the red line in Figure 1d). Therefore, the pathway from AB′ AFE to AC FE comprises two consecutive 90° switching events and long-distance interlayer sliding (Movie S1). This ferroelastic switching pathway resembles that of FE-order oxides, such as perovskite $BiFeO_3$ (71° followed by 109° switching)[42,48] and fluorite-structure $Hf_xZr_{1-x}O_2$ (90° switching)[7,49].

For the AB′ AFE to AB FE pathway (Movie S2 and pathway #2 in Figure 1c), the minimal distance between AB′ $(0.3a,0)$ and AB $(0.5a,0)$ stacking in the sliding map enables a direct 180° polarization switching. This occurs in alternating layers, where the polarization is oriented opposite to the applied electric field, resulting in an intermediate AB superlattice-like structure. The energy barrier of 180° switching pathway (blue line in Figure 1d) is much lower than that of the 90° pathway (1.2 vs. 2.6 meV/atom). However, the final AB FE phase exhibits relative energetic instability (Figure 1d), and can relax to the AC FE state upon removal of the electric field (Figure S5). The transformation from AB FE into the stable AC FE state is through another 90° switching without requiring high-barrier interlayer sliding (Supplementary Note 1).

To summarize, there are two energy-favorable switching pathways in SnSe from AFE to FE states (Figure 1f): (1) 90° pathway, involving two consecutive 90° switching events along the pathway #1 to AC FE state; and (2) 180° pathway, consisting of an initial top-layer switching and interlayer sliding to AB superlattice structure, followed by another top-layer switching to AB FE state (pathway #2). The 90° switching pathway exhibits a higher kinetic energy barrier (~2.5 meV/atom) yet leads to the thermodynamically preferred AC FE state. In contrast, the 180° switching pathway features a lower kinetic energy barrier (~1.2 meV/atom), but results in a metastable AB FE state with higher thermodynamic energy. The differences in thermodynamic energy between two FE phases (0.4 meV/atom) and energy barrier between the pathways (1.4 meV/atom) are both relatively small. This suggests that an applied electric field may lead to potential coexistence of both pathways during experimental switching events.

**Observation of *in-situ* switching at atomic scale**

To directly visualize the switching processes in SnSe, we employed STEM imaging to observe the atomic-scale structural evolution under *in-situ* biasing conditions. The SnSe flakes were synthesized by physical vapor deposition (PVD) and their square-shaped morphology is displayed in Figure S6. *In-situ* biasing experiment was performed using a micro-electromechanical systems (MEMS) chip-based holder. A focused ion beam (FIB) was used to transfer and thin the SnSe flakes (see Methods for details). Figure 2a and 2b illustrate the schematic of *in-situ* biasing setup and the corresponding scanning electron microscopy (SEM) image. To establish the in-plane biasing condition, Pt electrodes were deposited onto the SnSe sample using FIB. The central portion of the conductive Pt top layer was then cut off to fabricate an in-plane capacitor with an electrode spacing of approximately 5 μm. The ion milling induced damage only to the top ~100 nm surface and we specifically targeted the undamaged region beneath the surface-affected SnSe layer.

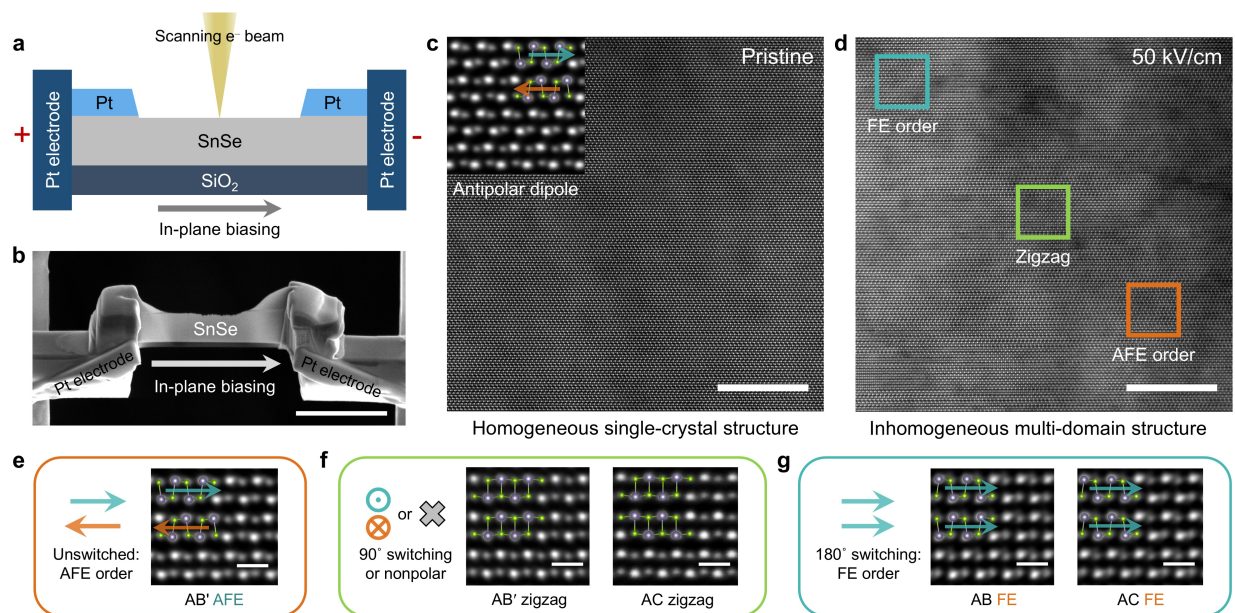

**Figure 2. Atomic-scale structural evolution of SnSe.** (a,b) Schematic of the *in-situ*, in-plane biasing of SnSe (a) and the corresponding SEM image of the transferred SnSe on MEMS chip (b). (c) Homogeneous single-crystal structure of pristine AFE-order AB′ stacking SnSe. (d) Inhomogeneous coherent multi-domain structure observed after applying 50 kV/cm biasing. The labeled square regions indicate three representative types of domains. (e) Magnified AFE domain, where slightly misaligned Sn-Sn ions imply pristine AB′ AFE structure. (f) Magnified zigzag domains, where the vertically aligned Sn-Se/Sn-Sn ions between two neighboring layers imply AB′/AC stacking, respectively. (g) Magnified FE domains, where the vertically aligned Sn-Sn/Sn-Se ions indicate AB/AC FE stacking, respectively. Scale bar: 3 μm in (b), 10 nm in (c,d), and 5 Å in (e-g).

Figure 2c and Figure S7a present high-angle annular dark-field (HAADF)-STEM images of the homogeneous, single-crystal SnSe sample on the biasing chip. The antiparallel Se shifts indicate AFE polarization order, with the interlayer structure consistent with AB′ stacking, as verified by large-scale polarization mapping (Figure S8). The HAADF-STEM image along the zigzag direction further confirms the pristine AB′ stacking order (Figure S9). Upon applying an increasing in-plane electric bias along the armchair direction, the antipolar dipoles remain stable below a critical threshold field, as confirmed by large-scale atomic structures and the corresponding fast Fourier transform (FFT) patterns (Figure S10). When the electric field reaches ~50 kV/cm, it triggers polarization switching. Owing to ferroelectric-ferroelastic coupling, the SnSe sample fractures at its thinnest central region to release lattice strain (Figure S11). X-ray energy dispersive spectroscopy (EDS) elemental mapping confirms that the SnSe chemical composition remains unaltered after switching (Figure S12).

Upon switching, the SnSe sample evolves into an inhomogeneous, yet lattice-coherent, multi-domain structure (Figure 2d). Owing to the rapid, kinetically driven switching process, both intermediate and fully switched FE states are observed, reflecting the non-uniformity of the final structure. The magnified atomic structure images show five representative domains: pristine AFE-order AB′ stacking domains (Figure 2e), AB′ zigzag domains (Figure 2f, left), AC zigzag domains (Figure 2f, right), AB FE (Figure 2g, left) and AC FE domains (Figure 2g, right). The coexistence of FE AB and FE AC domains suggest the coexistence of 90° and 180° switching pathways, balancing thermodynamic favorability (90° pathway) and minimized kinetic energy barrier (180° pathway) associated with interlayer sliding. In terms of the zigzag direction domains, although the atomic structures (Figure 2f) suggest multiple possible phases, the observed states most likely correspond to stable AFE and FE configurations within the energy landscape (Figure 1c and Figure S1). Therefore, by comparing these structures with DFT-predicted stable structures, the AB′ zigzag structures are identified as the AB′ AFE phase, an intermediate state in the 90° pathway. In addition, the AC zigzag structure is likely attributed to relaxation from AB FE structure in the 180° pathway after removing the applied bias, as predicted by DFT calculation (Figure S5 and Supplementary Note 1). These observations validate DFT-predicted states, offering experimental insights into the underlying switching pathways.

We further performed statistical analysis of these five domains over a >50 nm region where pronounced phase transitions occur (Figure S13 and Supplementary Note 2). The result indicates that the total area occupied by domains formed via the 90° switching pathway exceeds that of domains formed via the 180° pathway. This prevalence is attributed to the high applied electric field, which surpasses the energy barriers of both switching pathways, thereby promoting a thermodynamically driven preference for the lower-energy AC FE phase. In addition, the switched phases retain their configurations for a minimum of several hours, likely protected by domain-pinning effects, which kinetically inhibit the necessary interlayer sliding along all switching pathways. Such long-term stability surpasses that of traditional AFE perovskite oxides[50,51], owing to the minimal thermodynamic energy difference between AFE ground state and metastable FE phases in SnSe.

**Strain evolution and domain walls formation during switching**
The observed intermediate and switched states coexist in a layer-by-layer manner, as captured by the atomic-resolution images of FE-order and zigzag domains (Figure 3a,b). Such structures (schematically shown in Figure 3c) likely introduce lattice strain due to mismatched lattices between domains. As demonstrated by previous studies, reversible polarization switching is coupled with ferroelastic switching in few-layer MX flakes[33,52]. To quantitatively analyze the strain distribution in our sample, in-plane strain mapping was performed at each atomic position within the FE-order and zigzag domains (Figure 3d,e). In contrast to the nearly strain-free pristine state (Figure S7b), significant compressive strain is evident in both intermediate and final switched states. The histogram of in-plane strain distribution across the three states reveals a consistent compressive strain of approximately 6.5% induced by the switching processes (Figure 3f and Figure S14). This value closely matches the ferroelastic strain between the armchair and zigzag directions (~7%) (Table 1), suggesting that the strain arises from in-plane ferroelastic switching. When the structure reverts to armchair FE phases, the lattice retains the compressive strain to preserve lattice coherency across the multi-domain structure. Additionally, the significant in-plane strain inherited from the intermediate zigzag states stabilizes the final FE states, rendering the metastable AB and AC FE phases[38,41] non-volatile after the removal of the electric field. This stability is attributed to the smaller armchair lattice constant of the FE phases (Table S1), which prevents relaxation back to the AFE phase and would strongly modulate their electronic structures[53,54].

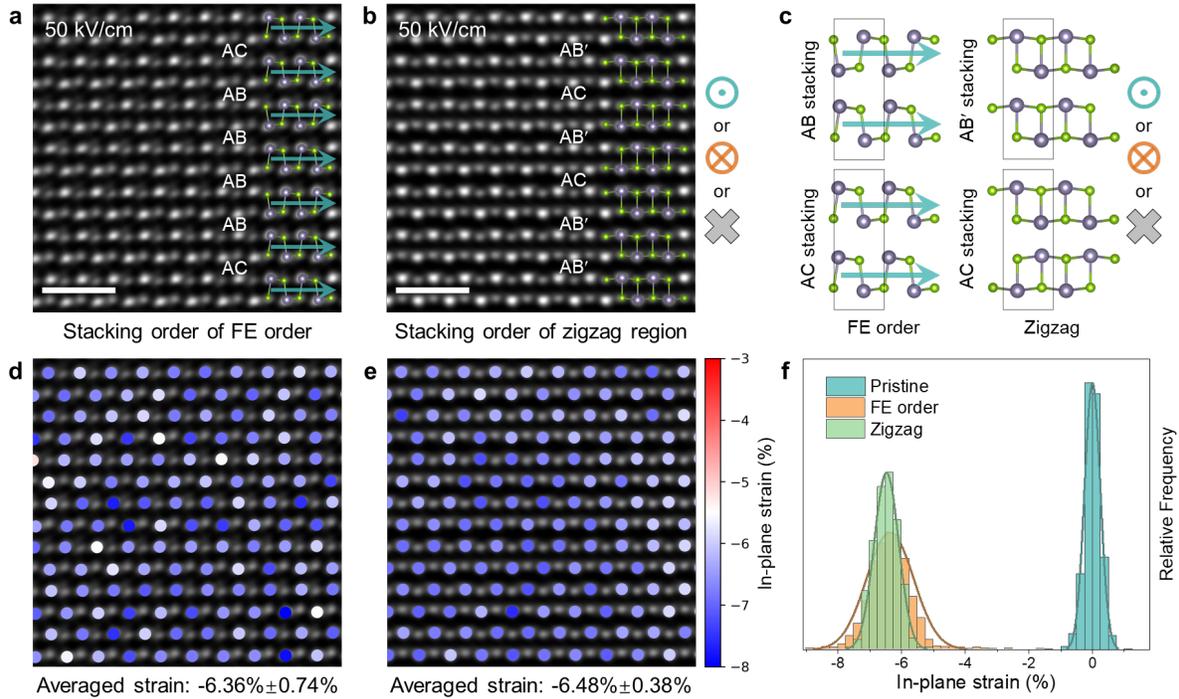

**Figure 3. Stacking structure and in-plane strain distribution in switched SnSe.** (a,b) HAADF-STEM images of FE-order (a) and zigzag (b) domains. (c) Atomic models of the observed FE-order and zigzag structure with AB, AB′, and AC stacking orders. (d,e) In-plane strain mapping corresponding to (a) and (b), respectively, where the pristine AB′ AFE structure used as the zero-strain reference. (f) Histogram of in-plane strain distribution in pristine and switched SnSe. Scale bar: 1 nm.

**Table 1. Measured in-plane lattice parameters and ferroelastic strain of pristine SnSe.**

|  | This work (Å) | Calculation (Å) | Ref. [55] (Å) |
| --- | --- | --- | --- |
| Armchair (a) | 4.47 | 4.49 | 4.44 |
| Zigzag (b) | 4.16 | 4.18 | 4.135 |
| Ferroelastic strain | -6.94% | -6.90% | -6.87% |

To investigate the domain wall structure after switching, we acquired HAADF-STEM images of the transition regions between adjacent domains within the inhomogeneous, coherent multi-domain area. Specifically, we analyzed the coherent intralayer interfaces between unswitched AB′ AFE and zigzag domains (Figure 4a), as well as between AB′ AFE and FE domains (Figure 4b). In-plane strain mapping of the domain wall regions reveals a strain gradient transitioning from AB′ AFE domains to zigzag or FE domains, forming a lattice-coherent interface with a high degree of crystallographic continuity (Figure 4c,d). The AB′ AFE domains exhibit minor local compressive strain, which progressively evolves to larger compressive strain in the zigzag and FE domains across the domain walls. The gradual interlayer sliding, coupled with concomitant in-plane strain evolution, facilitates the emergence of intermediate states at interface and polarization switching upon reaching a critical sliding displacement aligned with the switching pathway.

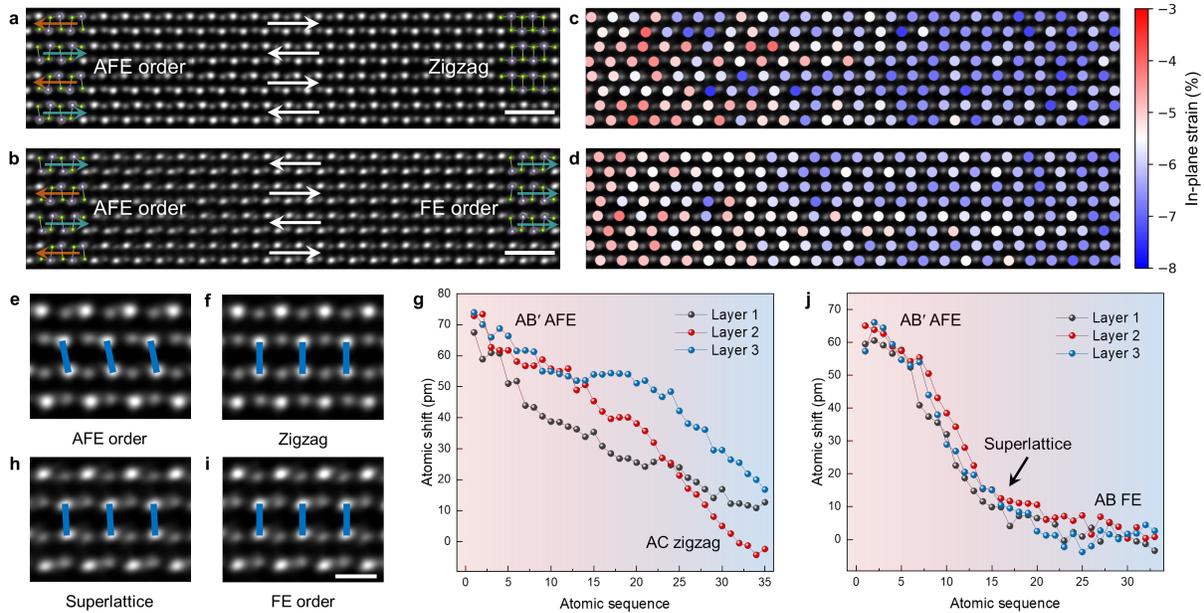

**Figure 4. Coherent domain wall structure of SnSe.** (a,b) HAADF-STEM images of an AFE-zigzag domain wall (a) and an AFE-FE domain wall (b). (c,d) In-plane strain mapping of (a) and (b), respectively, showing a strain gradient across the domain walls. The pristine AB′ AFE structure is used as the zero-strain reference. (e,f) Magnified HAADF-STEM images of AB′ AFE (e) and zigzag (f) structures, highlighting the interlayer Sn-Sn transition from a misaligned configuration (~70 pm atomic shift) to a vertically aligned zigzag domain. (g) Layer-by-layer interlayer atomic shift measurement from (a), revealing a smooth interlayer sliding transition. (h,i) Magnified ADF-STEM images of an intermediate superlattice structure captured at the domain wall in (b) (specific area shown in Fig. S8) and the FE-order structure, showing slight Sn-Sn misalignment (~10 pm atomic shift) in metastable superlattice configuration and the vertically aligned FE domain. (j) Layer-by-layer interlayer atomic shift measurement from (b), indicating a sharper and consistent sliding transition compared to (g). Scale bar: 1 nm in (a,b) and 4 Å in (i).

To better visualize interlayer sliding at these domain walls, we magnified the atomic structures in various regions of Figure 4a,b (specific area shown in Figure S15). Concerning the relative interlayer sliding between the AB′ AFE phase (Figure 4e) and the zigzag structure (Figure 4f), the interlayer Sn-Sn ions in the AB′ AFE phase exhibit a ~70 pm atomic misalignment, which gradually transitions to a vertically aligned configuration in the zigzag structure. Interlayer atomic shift measurements of Sn-Sn misalignment across the AFE-zigzag domain wall show a smooth transition with noticeable variations across layers due to the relaxation behavior of the zigzag

9 / 15

domain (Figure 4g and Supplementary Note). In contrast, the domain wall between AFE and FE regions exhibits more complex structures, such as the predicted unstable AB superlattice intermediate state in the 180° switching pathway. This state is characterized by a nonpolar-polar structure (Figure 1f) and resides within the domain wall region (Figure 4h). The AB superlattice state is also confirmed by the measured lattice parameters and periodic Sn-Se atomic distance within alternating polar and nonpolar layers (Table S2 and Figure S16). Compared to FE domains with vertically aligned Sn-Sn configurations (Figure 4i), the superlattice exhibits a ~10 pm atomic misalignment. Atomic shift measurements demonstrate a sharper and more consistent transition across layers, reflecting the completion of the 180° switching process. In essence, between AFE and FE domains with aligned armchair directions (e.g., Figure 4b), the domain walls exhibit a periodic charged configuration in the reversed polarization layers (tail-to-tail) (Figure S17)[56]. At the domain wall interface, the predicted AB superlattice structure is observed (Figure 1f), which does not relax into the AC FE zigzag structure (as shown in Figure 3b) due to the suppression of electrostatic energy arising from the charged boundary conditions.

## Discussion

In this study, we combined *in-situ* atomic-scale imaging with first-principles calculations to reveal a comprehensive understanding of energy-favorable switching pathways in SnSe. The uncovered intrinsic coupling between polarization switching, interlayer sliding, and lattice strain results in two coexisting switching pathways from AFE to FE phases. Given the ubiquitous interlayer vdW interactions in 2D ferroelectrics, our findings provide direct atomic-scale evidence of both the thermodynamic competition of FE phases with different stacking configurations and kinetic competition of switching pathways with different interlayer sliding distance. These insights into the underlying switching mechanisms offer a fundamental basis for potential MXs-based nanoelectronic applications, including non-volatile memories, logic transistors, nonlinear optoelectronic devices, and neuromorphic systems.

## Methods

### Material synthesis and transfer

Square-shaped SnSe flakes were synthesized on mica substrates via low-pressure PVD method with SnSe powder (99.999% metals basis, Thermo Scientific Chemicals) as precursor. The process was conducted at a base pressure of ~10 mTorr and a temperature of 440 °C, with a mixture of Ar/H$_2$ as the carrier gas, as reported in our previous work[32,38,39]. The SnSe samples were subsequently transferred onto silicon substrates using polymethyl methacrylate (PMMA), followed by sonication to detach the samples from the mica substrate.

### Sample fabrication and *in-situ* STEM imaging

A Protochips micro-electromechanical systems (MEMS) chip-based heating and biasing holder was used to achieve *in-situ* biasing condition. Using FEI Helios 660 focused ion beam (FIB), a cross-sectional lamella sample from a synthesized SnSe flake was transferred to the MEMS chip using an optimized Protochips FIB stub. During the thinning process, the accelerating voltage of the Ga ion beam gradually decreased in sequential steps (30 kV → 16 kV → 8 kV → 5 kV → 2 kV) to mitigate beam-induced amorphization of the specimen surface. In situ STEM imaging and X-ray EDS elemental mapping were performed using a double Cs-corrected FEI Titan Themis G3 (scanning) transmission electron microscope operated at 300 kV. HAADF-STEM images were acquired at each 1 V increment, ranging from 0 V to 26 V, with a collection semi-angle of 48–200 mrad. The applied electric field was calculated based on the applied voltage and the electrode gap (~5 μm).

### Quantitative image analysis

Fourier-filtered HAADF-STEM images were analyzed using CalAtom software to determine the precise atomic positions of Sn and Se ions by multiple-ellipse fitting[57]. The in-plane lattice parameters and strain $\varepsilon_{xx}$ were calculated by the atomic distance $L$ between adjacent brighter Sn ions within each unit cell (Figure S18):

$$\varepsilon_{xx} = \frac{L - L_0}{L_0}$$

where the reference $L_0$ for calculating the relative lattice strain is the averaged atomic distance along armchair or zigzag directions. Visualization of the atomic-scale strain mapping was performed using Python. The atomic shift between adjacent layers was quantified by measuring the Sn shift along the in-plane direction.

### First-principles calculations

First-principles study of vdW SnSe was performed using density functional theory (DFT) as implemented in the Vienna Ab initio Simulation Package (VASP)[58]. We adopted projector augmented wave method for treating core electrons[59], used the Perdew-Burke-Ernzerhof (PBE)-generalized gradient approximation (GGA) to compute exchange-correlation energy[60], and employed plane-wave basis with a cutoff energy of 500 eV and a Monkhorst-Pack k-point sampling grid[61] of 6×6×2. Crystal structures of vdW SnSe were optimized by relaxing the unit cell and all atomic positions, with the maximal residual atomic force of <0.02 eV Å$^{-1}$ and the total energy convergence criteria of <10$^{-5}$ eV. To properly take into account the weak van der Waals interaction, we adopted Grimme's DFT-D3 method with zero-damping function[62]. To understand the switching mechanism in vdW SnSe, we carried out potential energy surface calculations by first applying relative in-plane fractional shift between the two adjacent SnSe layers along their vdW plane on a grid of 11×11 from 0 to ½ *a*, and 0 to ½ *b*, then relaxing both lattice vector *c* and atomic coordinates of Sn and Se atoms along *c* while fixing lattice vectors *a* and *b* and the fractional atomic coordinates of Sn and Se atoms along *a* and *b*. To prevent the global drift, we further fix the position of one of the Sn atoms at the origin. The above potential energy surface calculations were performed for both AFE and FE ordered SnSe vdW layers, with AA, AB, AC, and AD stacking configurations located at (0,0), (½*a*, 0), (0, ½*b*), and (½*a*, ½*b*), respectively, as shown in Figure 1c. For the transformation pathway calculations, we used solid-state nudged elastic band (SS-NEB) method with climbing image[63]. Two transformation pathways were calculated with seven intermediate images as shown in Figure 1d: one pathway from the initial AB′ AFE structure to the intermediate 90° switched AB′ AFE (1a# green dashed arrows in Figure 1c) and then to the final AC FE structure through another 90° switching (1b# green dashed arrows in Figure 1c), and the other pathway from the initial AB′ AFE structure to the final AB FE structure through direct 180° switching (2# green dashed arrows in Figure 1c). The cell parameters and fractional atomic coordinates of all atoms in all seven intermediate images are fully optimized under the SS-NEB constraint.

**Acknowledgements**
X.L. acknowledges support from the Rice Advanced Materials Institute (RAMI) at Rice University as a RAMI Postdoctoral Fellow. X.L. and Y.H. acknowledge support from NSF (FUSE-2329111 and CMMI-2239545) and Welch Foundation (C-2065). We acknowledge the Electron Microscopy Center, Rice University. First-principles structural optimization and transition pathway by K.A. were supported by the Center for Reconfigurable Electronic Materials Inspired by Nonlinear Dynamics (reMIND), an Energy Frontier Research Center funded by the Department of Energy under award DE-SC0023353. S.K. acknowledges support by Texas A&M College of Engineering Horizons Fellowship. X.Q. acknowledges support from the Air Force Office of Scientific Research (AFOSR) under Grant No. FA9550-24-1-0207. Portions of this research were conducted with the advanced computing resources provided by Texas A&M High Performance Research Computing. S. H., X.L., and R.R. acknowledge support from Army Research Office and Army Research Laboratory via the Collaborative for Hierarchical Agile and Responsive Materials (CHARM) under cooperative agreement W911NF-19-2-0119. We acknowledge partial support from the U.S. Department of Energy, Office of Science, Office of Basic Energy Sciences, Materials Sciences and Engineering Division under Contract No. DE-AC02-05-CH11231 (Codesign of Ultra-Low-Voltage Beyond CMOS Microelectronics for the development of materials for low-power microelectronics). N.M., T. Z., and J. K. acknowledge support by U.S. Department of Energy, Office of Science, Basic Energy Sciences under Award No. DE-SC0020042 for the synthesis and transfer development of the SnSe samples.

**Author contributions:** Conceptualization: X.L. and Y.H. Methodology: X.L., K.A., R.R., J.K., X.Q., and Y. H. Investigation: X.L., K.A., C.S., N.M., S.K., K.Y., T.Z., S.H. Visualization: X.L., K.A., K.Y., and Y.H. Supervision: R.R., J.K., X.Q., and Y.H. Writing – original draft: X.L. and Y.H. Writing – review & editing: X.L., K.A., N.M., S.K., T.Z., S.H., R.R., J.K., X.Q., and Y. H. **Competing interests:** All authors declare that they have no competing interests. **Data availability:** All data needed to evaluate the conclusions in the paper are present in the paper and/or the Supplementary Materials.

# References

1. Valasek J. Piezo-Electric and Allied Phenomena in Rochelle Salt. *Phys. Rev.*, **17,** 475-481 (1921).
2. Junquera J, Nahas Y, Prokhorenko S, Bellaiche L, Íñiguez J, Schlom DG, *et al.* Topological phases in polar oxide nanostructures. *Rev. Mod. Phys.*, **95,** 025001 (2023).
3. Li L, Xie L, Pan X. Real-time studies of ferroelectric domain switching: a review. *Rep. Prog. Phys.*, **82,** 126502 (2019).
4. Das S, Tang YL, Hong Z, Gonçalves MAP, McCarter MR, Klewe C, *et al.* Observation of room-temperature polar skyrmions. *Nature*, **568,** 368-372 (2019).
5. Meier D, Selbach SM. Ferroelectric domain walls for nanotechnology. *Nat. Rev. Mater.*, **7,** 157-173 (2022).
6. Yang Q, Hu J, Fang Y-W, Jia Y, Yang R, Deng S, *et al.* Ferroelectricity in layered bismuth oxide down to 1 nanometer. *Science*, **379,** 1218-1224 (2023).
7. Li X, Liu Z, Gao A, Zhang Q, Zhong H, Meng F, *et al.* Ferroelastically protected reversible orthorhombic to monoclinic-like phase transition in $ZrO_2$ nanocrystals. *Nat. Mater.*, **23,** 1077-1084 (2024).
8. Vizner Stern M, Salleh Atri S, Ben Shalom M. Sliding van der Waals polytypes. *Nat. Rev. Phys.*, (2024).
9. Wu M, Li J. Sliding ferroelectricity in 2D van der Waals materials: Related physics and future opportunities. *Proc. Natl. Acad. Sci.*, **118,** e2115703118 (2021).
10. Cui C, Xue F, Hu W-J, Li L-J. Two-dimensional materials with piezoelectric and ferroelectric functionalities. *npj 2D Materials and Applications*, **2,** 18 (2018).
11. Wang C, You L, Cobden D, Wang J. Towards two-dimensional van der Waals ferroelectrics. *Nat. Mater.*, **22,** 542–552 (2023).
12. Zheng Z, Ma Q, Bi Z, de la Barrera S, Liu M-H, Mao N, *et al.* Unconventional ferroelectricity in moiré heterostructures. *Nature*, **588,** 71-76 (2020).
13. Jiang J, Xu L, Qiu C, Peng L-M. Ballistic two-dimensional InSe transistors. *Nature*, **616,** 470-475 (2023).
14. Yasuda K, Wang X, Watanabe K, Taniguchi T, Jarillo-Herrero P. Stacking-engineered ferroelectricity in bilayer boron nitride. *Science*, **372,** 1458-1462 (2021).
15. Stern MV, Waschitz Y, Cao W, Nevo, Watanabe K, Taniguchi T, *et al.* Interfacial ferroelectricity by van der Waals sliding. *Science*, **372,** 1462-1466 (2021).
16. Fei Z, Zhao W, Palomaki TA, Sun B, Miller MK, Zhao Z, *et al.* Ferroelectric switching of a two-dimensional metal. *Nature*, **560,** 336-339 (2018).
17. Wang X, Yasuda K, Zhang Y, Liu S, Watanabe K, Taniguchi T, *et al.* Interfacial ferroelectricity in rhombohedral-stacked bilayer transition metal dichalcogenides. *Nat. Nanotechnol.*, **17,** 367-371 (2022).
18. Weston A, Castanon EG, Enaldiev V, Ferreira F, Bhattacharjee S, Xu S, *et al.* Interfacial ferroelectricity in marginally twisted 2D semiconductors. *Nat. Nanotechnol.*, **17,** 390-395 (2022).
19. Rogée L, Wang L, Zhang Y, Cai S, Wang P, Chhowalla M, *et al.* Ferroelectricity in untwisted heterobilayers of transition metal dichalcogenides. *Science*, **376,** 973-978 (2022).
20. Deb S, Cao W, Raab N, Watanabe K, Taniguchi T, Goldstein M, *et al.* Cumulative polarization in conductive interfacial ferroelectrics. *Nature*, **612,** 465-469 (2022).
21. Ding W, Zhu J, Wang Z, Gao Y, Xiao D, Gu Y, *et al.* Prediction of intrinsic two-dimensional ferroelectrics in $In_2Se_3$ and other $III_2$-$VI_3$ van der Waals materials. *Nat. Commun.*, **8,** 14956 (2017).
22. Modi G, Parate SK, Kwon C, Meng AC, Khandelwal U, Tullibilli A, *et al.* Electrically driven long-range solid-state amorphization in ferroic $In_2Se_3$. *Nature*, (2024).
23. Han W, Zheng XD, Yang K, Tsang CS, Zheng FY, Wong LW, *et al.* Phase-controllable large-area two-dimensional $In_2Se_3$ and ferroelectric heterophase junction. *Nat. Nanotechnol.*, **18,** 55-63 (2023).
24. Liu F, You L, Seyler KL, Li X, Yu P, Lin J, *et al.* Room-temperature ferroelectricity in $CuInP_2S_6$ ultrathin flakes. *Nat. Commun.*, **7,** 12357 (2016).

25. Brehm JA, Neumayer SM, Tao L, O'Hara A, Chyasnavichus M, Susner MA, *et al.* Tunable quadruple-well ferroelectric van der Waals crystals. *Nat. Mater.*, **19,** 43-48 (2020).
26. Gou J, Bai H, Zhang XL, Huang YL, Duan SS, Ariando A, *et al.* Two-dimensional ferroelectricity in a single-element bismuth monolayer. *Nature*, **617,** 67-72 (2023).
27. Song Q, Occhialini CA, Ergeçen E, Ilyas B, Amoroso D, Barone P, *et al.* Evidence for a single-layer van der Waals multiferroic. *Nature*, **602,** 601-605 (2022).
28. Chang K, Liu J, Lin H, Wang N, Zhao K, Zhang A, *et al.* Discovery of robust in-plane ferroelectricity in atomic-thick SnTe. *Science*, **353,** 274-278 (2016).
29. Barraza-Lopez S, Fregoso BM, Villanova JW, Parkin SSP, Chang K. Colloquium: Physical properties of group-IV monochalcogenide monolayers. *Rev. Mod. Phys.*, **93,** 011001 (2021).
30. Wang H, Qian X. Giant Optical Second Harmonic Generation in Two-Dimensional Multiferroics. *Nano Lett.*, **17,** 5027-5034 (2017).
31. Wang H, Qian X. Two-dimensional multiferroics in monolayer group IV monochalcogenides. *2d Mater.*, **4,** 015042 (2017).
32. Chiu M-H, Ji X, Zhang T, Mao N, Luo Y, Shi C, *et al.* Growth of Large-Sized 2D Ultrathin SnSe Crystals with In-Plane Ferroelectricity. *Adv. Electron. Mater.*, **9,** 2201031 (2023).
33. Luo Y, Mao N, Ding D, Chiu M-H, Ji X, Watanabe K, *et al.* Electrically switchable anisotropic polariton propagation in a ferroelectric van der Waals semiconductor. *Nat. Nanotechnol.*, **607,** 1-7 (2023).
34. Guan Z, Zheng YZ, Tong WY, Zhong N, Cheng Y, Xiang PH, *et al.* 2D Janus Polarization Functioned by Mechanical Force. *Adv. Mater.*, **36,** 2403929 (2024).
35. Nanae R, Kitamura S, Chang Y-R, Kanahashi K, Nishimura T, Moqbel R, *et al.* Bulk Photovoltaic Effect in Single Ferroelectric Domain of SnS Crystal and Control of Local Polarization by Strain. *Adv. Funct. Mater.*, **34,** 2406140 (2024).
36. Wang H, Qian X. Ferroicity-driven nonlinear photocurrent switching in time-reversal invariant ferroic materials. *Science Advances*, **5,** eaav9743 (2019).
37. Hanakata PZ, Carvalho A, Campbell DK, Park HS. Polarization and valley switching in monolayer group-IV monochalcogenides. *Phys. Rev. B*, **94,** 035304 (2016).
38. Shi C, Mao N, Zhang K, Zhang T, Chiu M-H, Ashen K, *et al.* Domain-dependent strain and stacking in two-dimensional van der Waals ferroelectrics. *Nat. Commun.*, **14,** 7168 (2023).
39. Mao N, Luo Y, Chiu MH, Shi C, Ji X, Pieshkov TS, *et al.* Giant Nonlinear Optical Response via Coherent Stacking of In-Plane Ferroelectric Layers. *Adv. Mater.*, **35,** e2210894 (2023).
40. Chang YR, Nanae R, Kitamura S, Nishimura T, Wang HN, Xiang YB, *et al.* Shift-Current Photovoltaics Based on a Non-Centrosymmetric Phase in In-Plane Ferroelectric SnS. *Adv. Mater.*, **35,** 2301172 (2023).
41. Xu B, Deng J, Ding X, Sun J, Liu JZ. Van der Waals force-induced intralayer ferroelectric-to-antiferroelectric transition via interlayer sliding in bilayer group-IV monochalcogenides. *npj Comput. Mater.*, **8,** 47 (2022).
42. Caretta L, Shao Y-T, Yu J, Mei AB, Grosso BF, Dai C, *et al.* Non-volatile electric-field control of inversion symmetry. *Nat. Mater.*, **22,** 207-215 (2023).
43. Wang H, Qian X. Ferroelectric nonlinear anomalous Hall effect in few-layer $WTe_2$. *npj Comput. Mater.*, **5,** 119 (2019).
44. Xiao J, Wang Y, Wang H, Pemmaraju CD, Wang S, Muscher P, *et al.* Berry curvature memory through electrically driven stacking transitions. *Nat. Phys.*, **16,** 1028-1034 (2020).
45. Kwon KC, Zhang YS, Wang L, Yu W, Wang XJ, Park IH, *et al.* In-Plane Ferroelectric Tin Monosulfide and Its Application in a Ferroelectric Analog Synaptic Device. *ACS Nano*, **14,** 7628-7638 (2020).
46. Guan Z, Zhao Y, Wang X, Zhong N, Deng X, Zheng Y, *et al.* Electric-Field-Induced Room-Temperature Antiferroelectric–Ferroelectric Phase Transition in van der Waals Layered GeSe. *ACS Nano*, **16,** 1308-1317 (2022).

47. Wang C, Li Z, Cheng Y, Weng X-J, Bu Y, Zhai K, *et al.* Reversible shuffle twinning yields anisotropic tensile superelasticity in ceramic GeSe. *Nat. Nanotechnol.*, (2025).
48. Heron JT, Bosse JL, He Q, Gao Y, Trassin M, Ye L, *et al.* Deterministic switching of ferromagnetism at room temperature using an electric field. *Nature*, **516,** 370-373 (2014).
49. Ding W, Zhang Y, Tao L, Yang Q, Zhou Y. The atomic-scale domain wall structure and motion in $HfO_2$-based ferroelectrics: A first-principle study. *Acta Mater.*, **196,** 556–564 (2020).
50. Zeng F, Zeng H, Zhang Y, Shen M, Hu Y, Gao S, *et al.* Ultrahigh Energy Storage in (Ag,Sm)(Nb,Ta)O3 Ceramics with a Stable Antiferroelectric Phase, Low Domain-Switching Barriers, and a High Breakdown Strength. *ACS Applied Materials & Interfaces*, **16,** 51170-51181 (2024).
51. Nadaud K, Borderon C, Renoud R, Bah M, Ginestar S, Gundel HW. Study of the long time relaxation of the weak ferroelectricity in PbZrO3 antiferroelectric thin film using Positive Up Negative Down and First Order Reversal Curves measurements. *Thin Solid Films*, **773,** 139817 (2023).
52. Brillson LJ, Burstein E, Muldawer L. RAMAN OBSERVATION OF FERROELECTRIC PHASE-TRANSITION IN SNTE. *Phys. Rev. B*, **9,** 1547-1551 (1974).
53. Jaikissoon M, Köroğlu Ç, Yang JA, Neilson K, Saraswat KC, Pop E. CMOS-compatible strain engineering for monolayer semiconductor transistors. *Nat. Electron.*, **7,** 885-891 (2024).
54. Peng Z, Chen X, Fan Y, Srolovitz DJ, Lei D. Strain engineering of 2D semiconductors and graphene: from strain fields to band-structure tuning and photonic applications. *Light Sci. Appl.*, **9,** 190 (2020).
55. Zhao L-D, Lo S-H, Zhang Y, Sun H, Tan G, Uher C, *et al.* Ultralow thermal conductivity and high thermoelectric figure of merit in SnSe crystals. *Nature*, **508,** 373-377 (2014).
56. Cheng M, Si Y, Li N, Guan J. Synergy of Charged Domain Walls in 2D In-Plane Polarized Ferroelectric GeS for Photocatalytic Water Splitting. *J. Am. Chem. Soc.*, **146,** 26567-26573 (2024).
57. Zhang Q, Zhang LY, Jin CH, Wang YM, Lin F. CalAtom: A software for quantitatively analysing atomic columns in a transmission electron microscope image. *Ultramicroscopy*, **202,** 114-120 (2019).
58. Kresse G, Furthmüller J. Efficient iterative schemes for ab initio total-energy calculations using a plane-wave basis set. *Phys. Rev. B*, **54,** 11169-11186 (1996).
59. Blöchl PE. Projector augmented-wave method. *Phys. Rev. B*, **50,** 17953-17979 (1994).
60. Perdew JP, Burke K, Ernzerhof M. Generalized Gradient Approximation Made Simple. *Phys. Rev. Lett.*, **77,** 3865-3868 (1996).
61. Monkhorst HJ, Pack JD. Special points for Brillouin-zone integrations. *Phys. Rev. B*, **13,** 5188 (1976).
62. Grimme S, Antony J, Ehrlich S, Krieg H. A consistent and accurate ab initio parametrization of density functional dispersion correction (DFT-D) for the 94 elements H-Pu. *J. Chem. Phys.*, **132,** 154104 (2010).
63. Henkelman G, Uberuaga BP, Jónsson H. A climbing image nudged elastic band method for finding saddle points and minimum energy paths. *The Journal of Chemical Physics*, **113,** 9901-9904 (2000).

# Supplementary Materials

**Table of contents:**

    Supplementary Notes 1–3

    Figures S1–S18

    Tables S1 and S2

    Captions of Movies S1 and S2

## Supplementary Note 1
### Decoupling interlayer sliding and intralayer polarization switching in switching pathways

In 2D multilayer ferroelectric systems, polarization switching is strongly coupled with interlayer sliding[15,19,20,41], which complicates the description of switching pathways, especially in SnSe. To simplify the description, we decouple relative interlayer sliding and intralayer polarization switching, treating them as two separate aspects within the switching pathways. First, we denote the interlayer sliding direction and distance using coordinates on sliding maps (Figure 1c,e and Figure S1). In terms of intralayer polarization switching, this process involves two main components: the relative displacement between Sn and Se ions (indicative of ferroelectric switching) and the corresponding strain relaxation. Strain relaxation can arise either from 90° ferroelastic strain (interchange between long armchair and short zigzag axes) or from lattice difference between relaxed structures (Table S1). The armchair-zigzag interchange corresponds to a mirror symmetry in the energy landscape along the diagonal, as shown in Figure S1.

Notably, the stacking structure can transform between AB and AC through a single 90° switching event without interlayer sliding (see Figure below) due to the definitions of AB and AC stackings. Thus, AB FE and AC FE can readily transform to each other by 90° switching with a lower energy barrier (Figure S5), explaining our observation of relaxed AC FE zigzag structures stabilized from the unstable AB superlattice configuration. However, when aligned with an external electric field for a 180° switching event, the system requires maximum sliding from $(0.5a,0)$ to $(0,0.5b)$ and an additional 90° switching event. In summary, the green lines in Figure 1c and Figure S1 indicate only the interlayer sliding direction and distance along different pathways. But both interlayer sliding and intralayer polarization switching are included in the switching pathway calculation to obtain more accurate results.

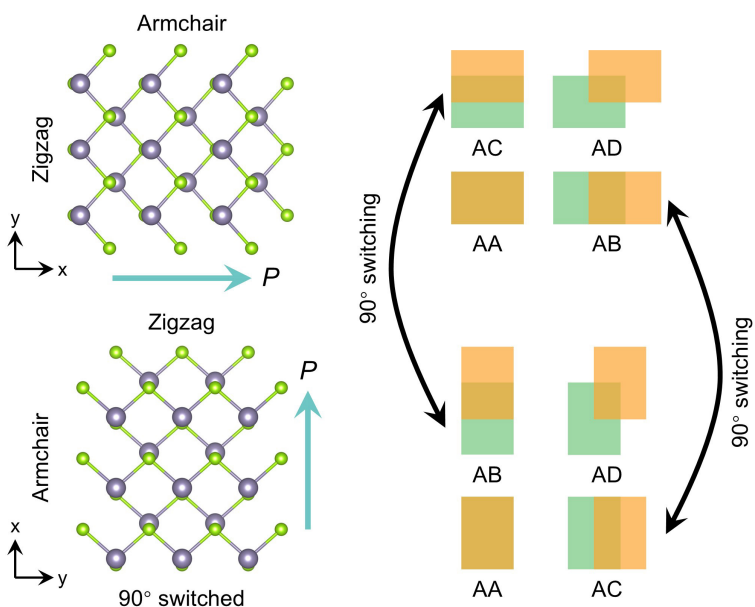

**AB and AC stacking configurations can transform into each other through a single 90° switching event without requiring interlayer sliding.**

## Supplementary Note 2
**Statistical analysis of the domain structure after switching**

Consistent with our discussion in Figure 2e–g, there are five types of domains in the switched SnSe, which are mapped in Figure S13. Statistical analysis of the domain areas reveals that the AB′ AFE domain accounts for over 50% of the total area, corresponding to the unswitched regions. AB′ zigzag and AC FE domains correspond to the intermediate and final state in 90° switching pathway, respectively (as discussed in Figure 1). The summed area fraction for domains associated with this 90° switching pathway is approximately 40%, indicating that 90° switching is the dominant mechanism in the switching process.

The AB FE domain corresponds to the final state of the 180° switching pathway. However, upon removal of the electric field, the AB FE state likely relaxes into a more stable AC FE configuration, with its polarization oriented perpendicular to that of the final AB FE state and without requiring interlayer sliding, as discussed on Page 6 of the main text. In the phase map (Figure S13), the AC zigzag denotes an AC FE structure with polarization perpendicular to the imaging plane. Therefore, the total area fraction associated with the 180° switching pathway (AC zigzag + AB FE) is approximately 10%. These results suggest that the 180° switching pathway is less favorable, leading to metastable final FE states that readily relax into a polarization orientation perpendicular to the applied electric field.

The relaxation from AB FE to AC zigzag structure is verified using solid-state nudged elastic band (SS-NEB) method (Figure S5). As discussed in Supplementary Note 1 and main text, 90° switching from AB FE to AC zigzag doesn't require interlayer sliding. Consequently, only a very small energy barrier (~0.4 meV/atom) is required for the relaxation, explaining the experimentally observed large-area relaxation to AC zigzag domains.

In summary, although the applied high electric field surpasses the switching barriers and induces coexistence of both pathways, the thermodynamically favored process ultimately leads to a preference for the 90° pathway. Notably, since determination of the stacking order requires at least two adjacent layers, some layers near vertical domain boundaries may have been double-counted in the total area calculation; however, this does not affect the overall conclusions drawn from the statistical analysis.

**Supplementary Note 3**
**Mechanical stability limitation in *in-situ* STEM imaging experiments**
As demonstrated by previous studies[33,46], electric-field-driven AFE-to-FE phase transition and reversible polarization switching constitute intrinsic properties of group-IV MXs. In our *in-situ* STEM experiments, an electron-transparent cross-sectional SnSe lamella was prepared for atomic-scale observation, which exhibited compromised mechanical stability, resulting in localized cracking after applying a large electric field. While the mechanical stability of the sample varies between SnSe flake and lamella, the intrinsic energy landscapes governing phase transitions are universal across both morphologies. AFE-to-FE switching invariably follows energetically favorable pathways dictated by energy landscapes. Consequently, the switching mechanisms observed experimentally and modeled theoretically are applicable to van der Waals MX material systems.

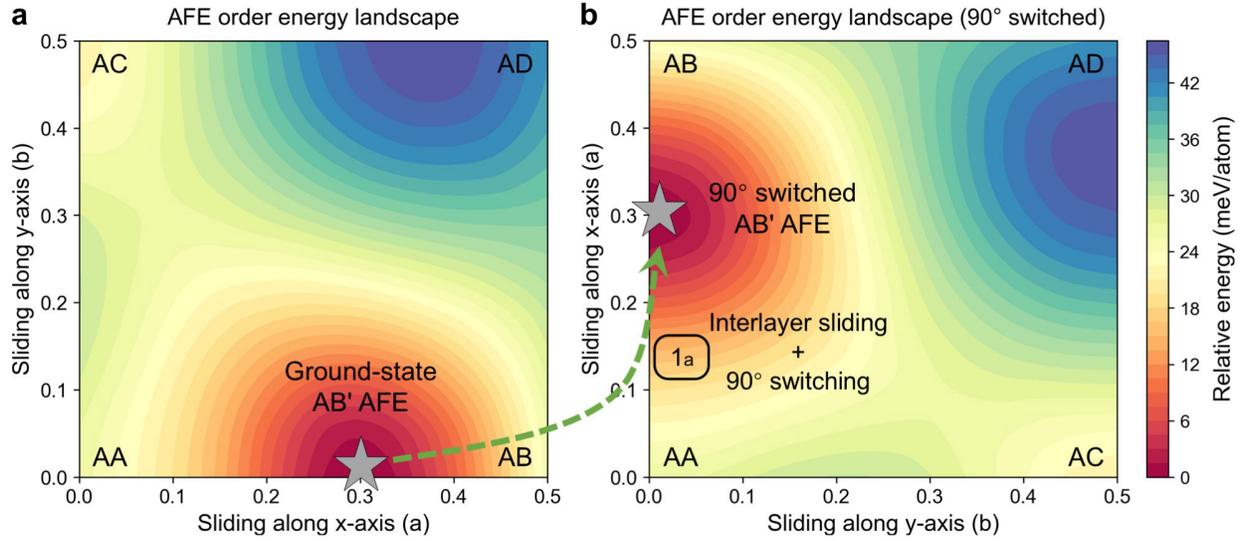

**Figure S1. Schematic of 1a switching pathway in the energy landscape.** (a) Energy landscape of AFE-order SnSe. (b) After 90° switching, the *x* and *y* axes (armchair and zigzag) are interchanged, resulting in the energy landscape mirrored by the diagonal line. The switching process encompasses both interlayer sliding from (0.3*a*, 0) to (0, 0.3*b*) and intralayer 90° polarization switching through relative shifts of Se atoms with respect to Sn atoms.

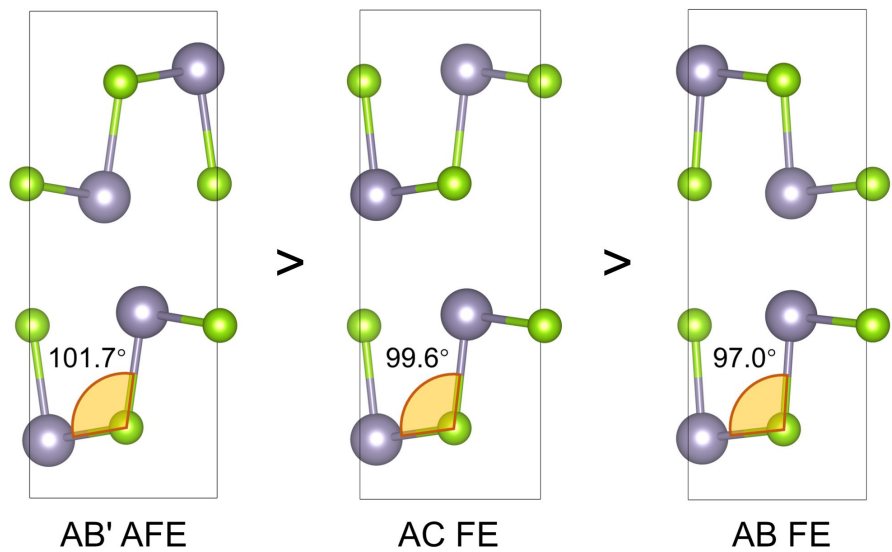

**Figure S2. Comparison of electric polarization between three energy-favorable states.**

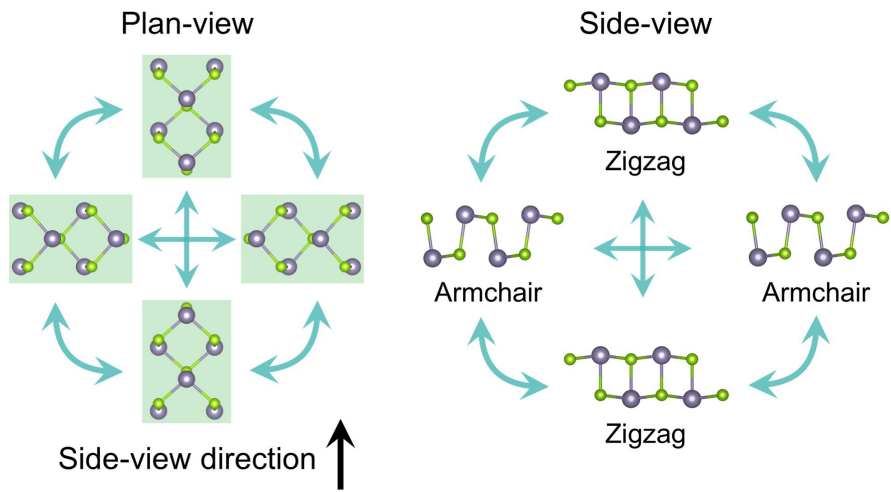

**Figure S3. Schematics of plan-view 90° and 180° intralayer polarization switching and the side-view armchair-to-zigzag 90° switching.**

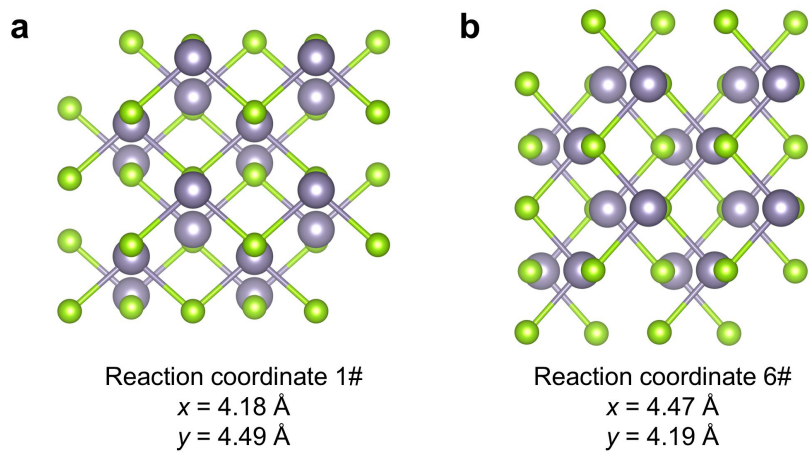

**Figure S4. Calculated atomic structures of AFE-order SnSe along the 90° switching pathway.** (a) Initial structure. (b) Intermediate 90° switched-like structure.

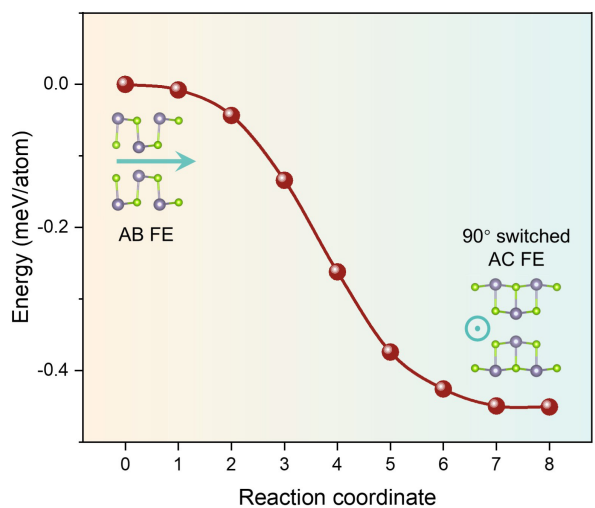

**Figure S5. Small switching barrier of 90° switching from AB FE to AC FE without interlayer sliding.**

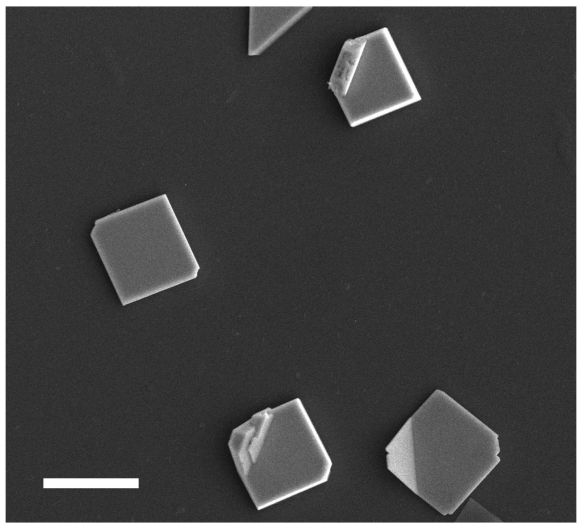

**Figure S6. SEM image of as-synthesized square-shaped SnSe flakes. Scale bar: 5 µm.**

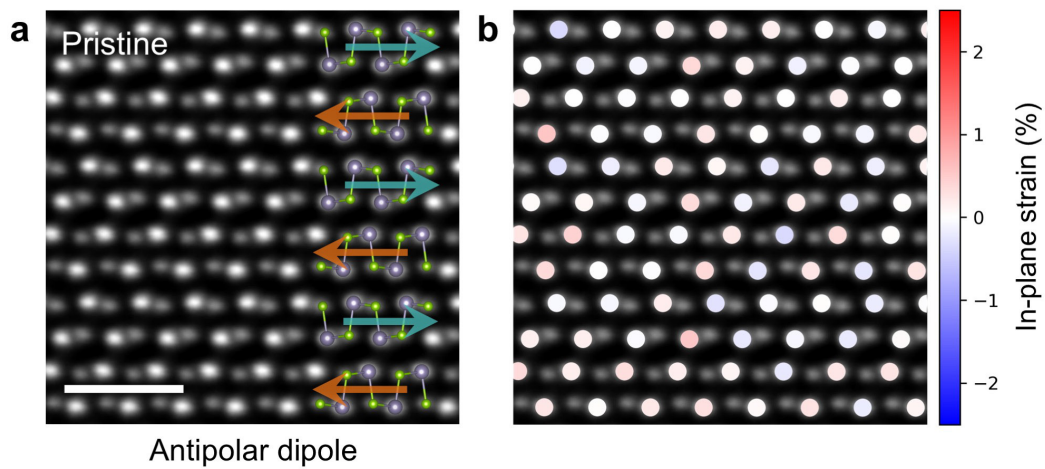

Antipolar dipole

**Figure S7. Antipolar structure and in-plane strain mapping of SnSe.** (a) Pristine SnSe along armchair direction shows ground-state AB stacking order with AFE polarization order. Scale bar: 1 nm. (b) The corresponding in-plane strain mapping shows ~0% intralayer strain of pristine structure.

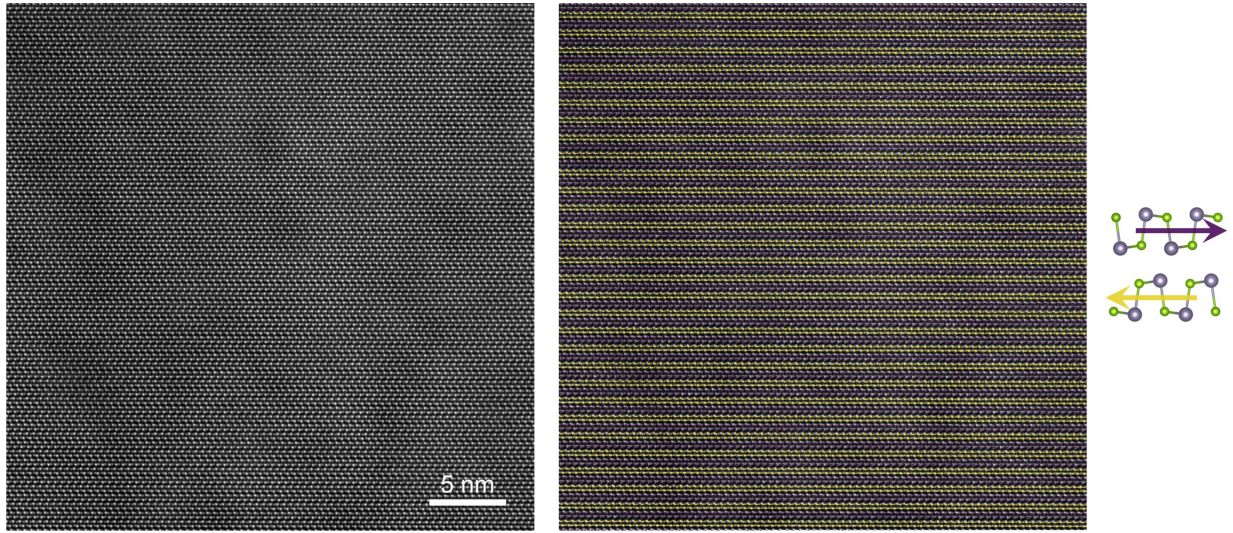

**Figure S8.** Large-scale atomic structure and the corresponding polarization mapping show homogeneous antipolar order in the pristine SnSe.

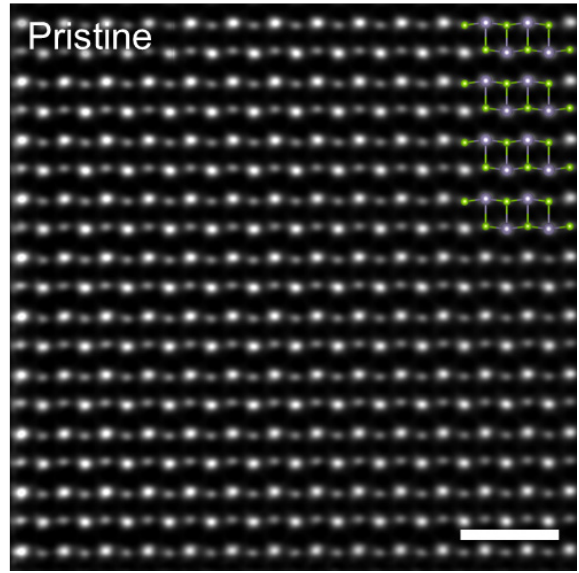

**Figure S9. Pristine SnSe along zigzag direction shows ground-state AB stacking structure.** Scale bar: 1 nm.

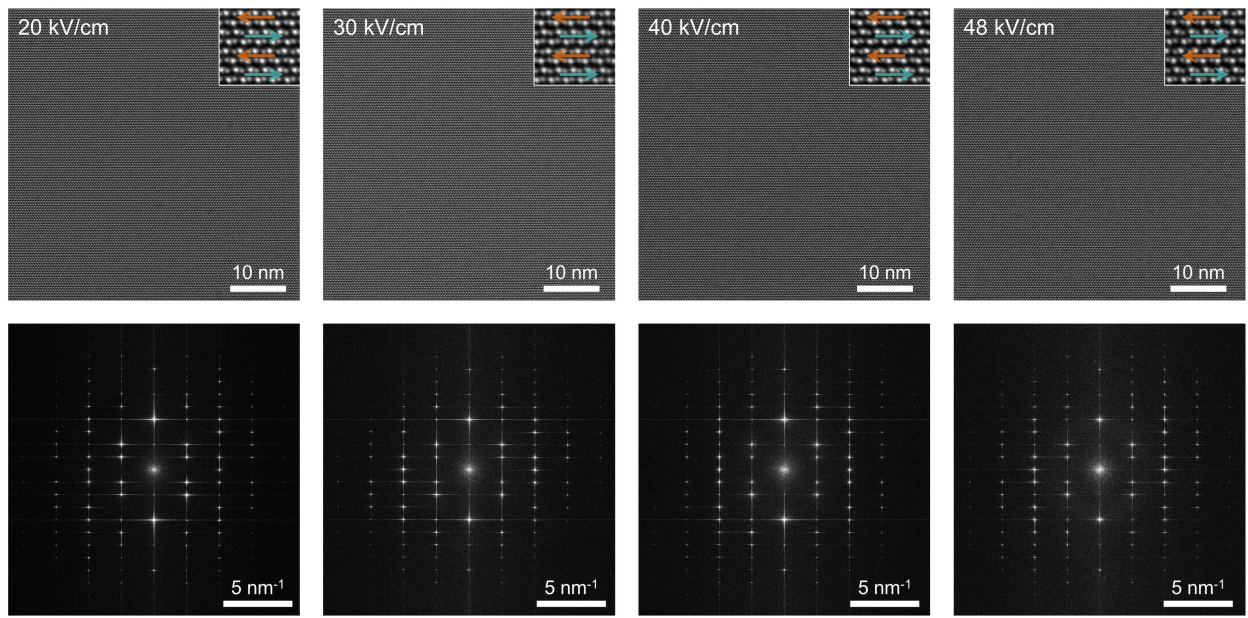

**Figure S10.** Low-magnification HAADF-STEM images and the corresponding FFT patterns under different electric fields.

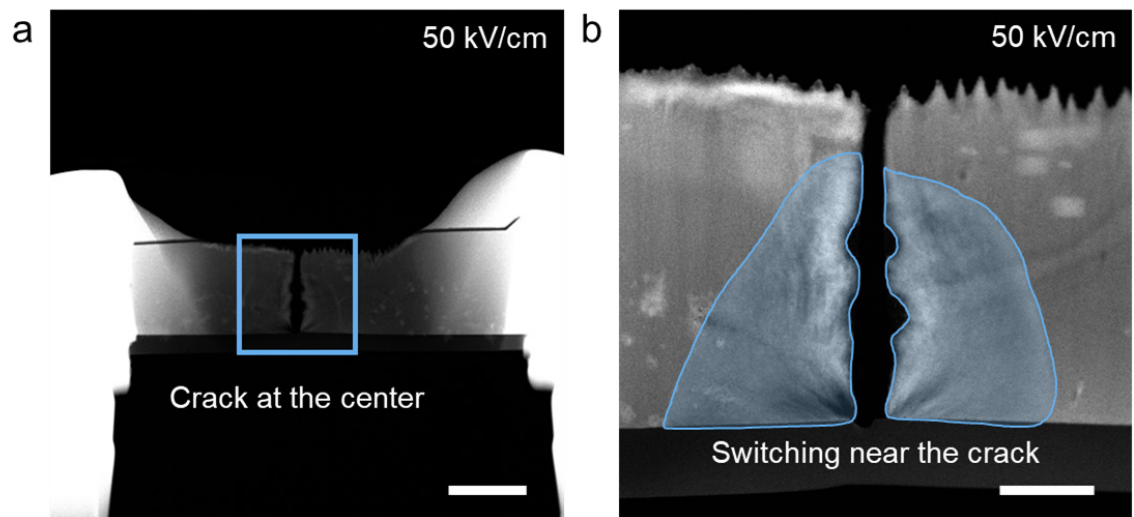

**Figure S11. Morphology of SnSe lamella after in-situ switching.** (a) The SnSe crack appears in the thinnest central part. (b) Switching occurs in the bending region near the crack. Scale bar: 1 μm in (a) and 300 nm in (b).

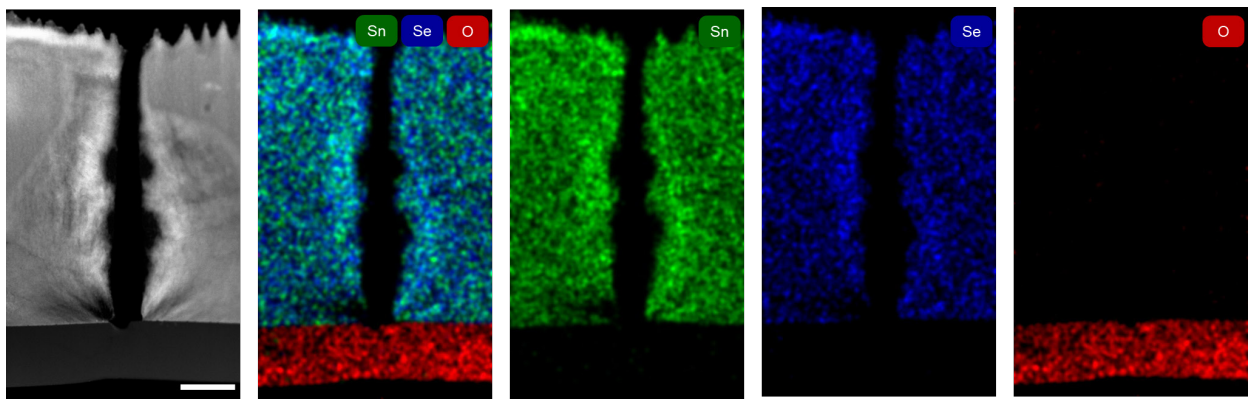

**Figure S12. Elemental mapping after switching.** HAADF-STEM image and the corresponding X-ray EDS elemental mapping show that the chemical component does not change after switching. The bottom layer is SiO$_2$. Scale bar: 200 nm.

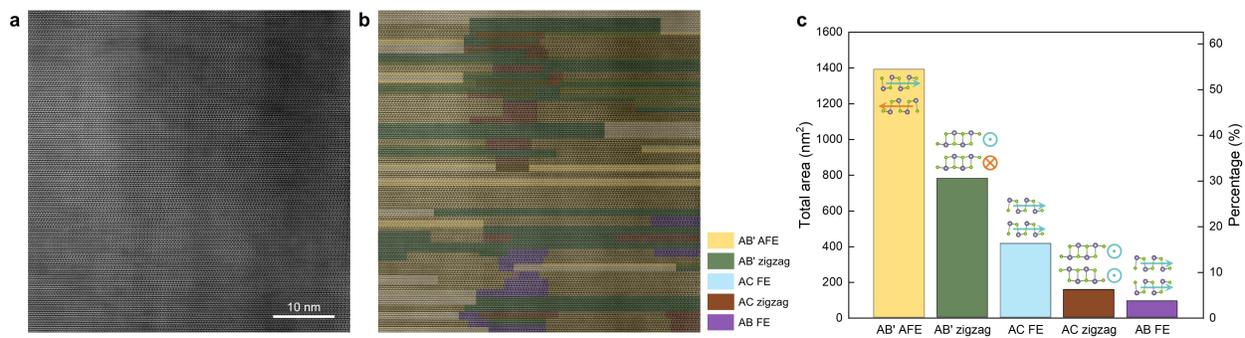

**Figure S13. Large-scale statistical analysis of domain structures in switched SnSe.** (a) Atomic structure after switching. (b) The corresponding phase mapping. (c) The statistical analysis of different domain structures.

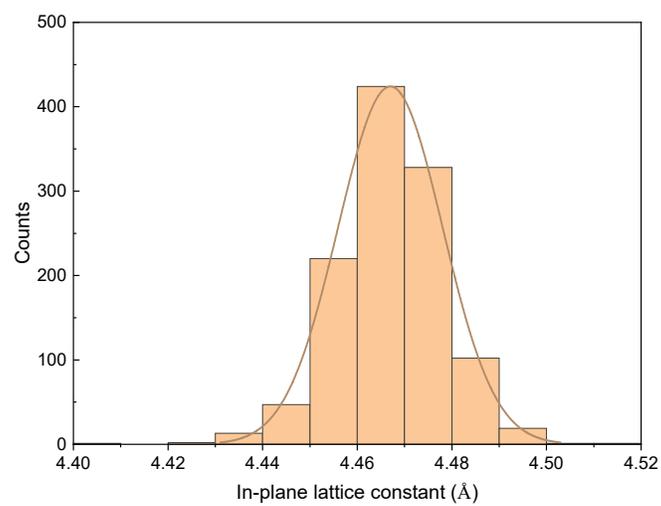

**Figure S14. Histogram of pristine in-plane lattice constant shows 4.47 ± 0.01 Å accuracy of atomic measurement.**

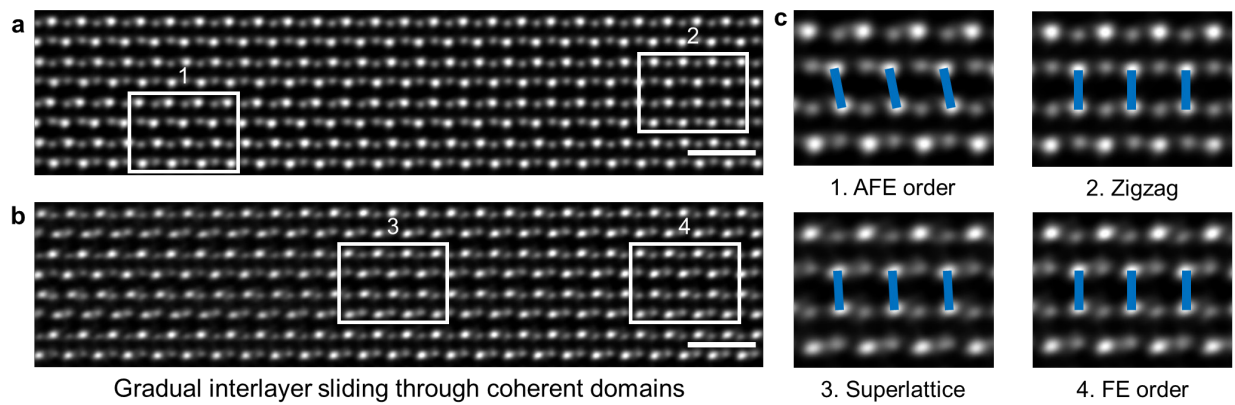

**Figure S15. Sampling regions in Figure 4.** Magnified regions in Figure 4a,b (a,b). The magnified region is labeled in (c).

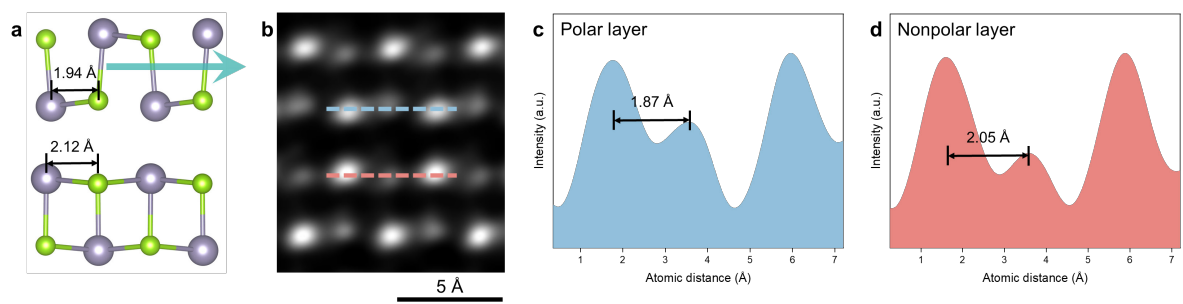

**Figure S16. Periodic Sn-Se atomic distance in AB superlattice state.** (a,b) Atomic schematic (a) and HAADF-STEM images of AB superlattice state. (c,d) Line profiles in (b) measure the Sn-Se distance in polar and nonpolar layer.

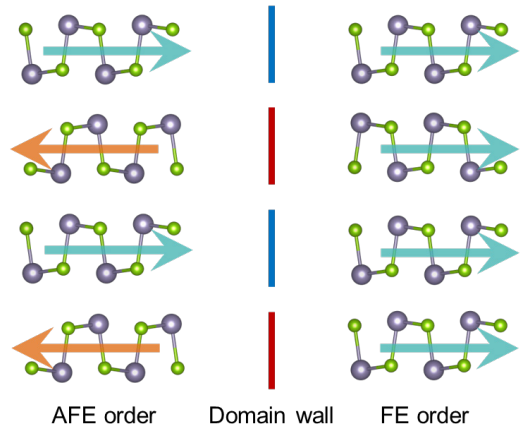

AFE order    Domain wall    FE order

**Figure S17. Observed periodically charged domain walls between AFE and FE order domains.** Red lines represent tail-to-tail charged domain walls and blue lines represent head-to-tail neutral domain walls.

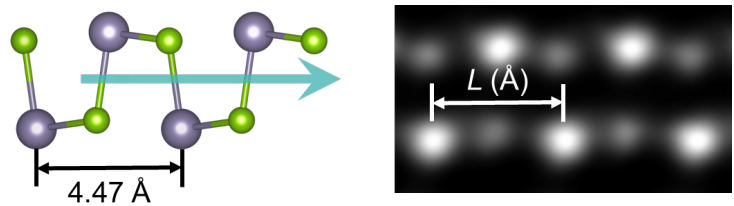

**Figure S18. Schematic for atomic distance calculation.**

**Table S1. Stable and metastable states of multilayer SnSe.**

| Polarization order | Stacking structure | a (Å) | b (Å) | $\varepsilon_{xx}$ | Energy (meV/atom) |
|:---:|:---:|:---:|:---:|:---:|:---:|
| AFE | AB′ | 4.49 | 4.18 | 0 | 0 |
| FE | AC | 4.36 | 4.23 | -2.90% | 0.8 |
| FE | AB | 4.31 | 4.23 | -4.01% | 1.2 |

**Table S2. Comparation between experimental measurement and DFT calculations of AB superlattice state.**

|  | Experiment (Å) | Calculation (Å) |
|:---:|:---:|:---:|
| In-plane (a) | 4.22 | 4.30 |
| Out-of-plane (c) | 11.85 | 11.86 |

**Movie S1.** DFT-calculated 90° switching pathway illustrated from both plan-view and side-view perspectives.

**Movie S2.** DFT-calculated 180° switching pathway illustrated from both plan-view and side-view perspectives.